\documentclass[12pt,pre,aps,showpacs,preprint]{revtex4}

\usepackage{graphics}
\usepackage{graphicx}

\begin{document}

\title{Dense Packings of Superdisks and the Role of Symmetry}

\author{Y. Jiao}


\affiliation{Department of Mechanical and Aerospace Engineering, 
Princeton University, Princeton New Jersey 08544, USA}

\author{F. H. Stillinger}


\affiliation{Department of Chemistry, Princeton University,
Princeton New Jersey 08544, USA}

\author{S. Torquato}

\email{torquato@electron.princeton.edu}

\affiliation{Department of Chemistry, Princeton University,
Princeton New Jersey 08544, USA}

\affiliation{Princeton Institute for the Science and Technology of
Materials, Princeton University, Princeton New Jersey 08544, USA}

\affiliation{Program in Applied and Computational Mathematics, 
Princeton University, Princeton New Jersey 08544, USA}

\affiliation{Princeton Center for Theoretical Physics, Princeton
University, Princeton New Jersey 08544, USA}

\begin{abstract}
We construct the densest known two-dimensional packings of superdisks 
in the plane whose shapes are defined by $|x|^{2p}+|y|^{2p}\le1$,
which contains both convex-shaped particles ($p\ge 0.5$, with the
circular-disk case $p=1$) and concave-shaped particles ($0<p<0.5$). 
The packings of the convex cases with $p \ge 1$ generated by a
recently developed event-driven molecular dynamics (MD) simulation
algorithm [Donev, Torquato and Stillinger, J. Comput. Phys. {\bf
202} (2005) 737] suggest exact constructions of the densest known packings. 
We find that the packing density (covering
fraction of the particles) $\phi$ increases dramatically as the
particle shape moves away from the "circular-disk" point ($p=1$).
In particular, we find that the maximal packing densities of
superdisks for certain $p \neq 1$ are achieved by one of the two
families of Bravais lattice packings, which provides additional
numerical evidence for Minkowski's conjecture concerning the
critical determinant of the region occupied by a superdisk.
Moreover, our analysis on the generated packings reveals that the
broken rotational symmetry of superdisks influences the packing
characteristics in a non-trivial way. We also propose an
analytical method to construct dense packings of concave
superdisks based on our observations of the structural properties
of packings of convex superdisks.
\end{abstract}


\maketitle


Packing problems, such as how densely given solid objects can fill
$d$-dimensional Euclidean space $\Re^d$, have been a source of
fascination to mathematicians and scientists for centuries, and
continue to intrigue them today. A basic characteristic of a
packing (a large collection of nonoverlapping particles) is the
packing density $\phi$, defined as the fraction of space covered
by the particles. Estimation of the maximal packing density (the
maximal fraction of space covered by the particles) $\phi_{max}$
of a given nonoverlapping body arranged on the sites of a Bravais
lattice (i.e., a Bravais lattice packing) is one of the basic
problems in the geometry of numbers \cite{Cassel, Conway}. Dense
packings of nonoverlapping (hard) particles have served as useful
models to understand the structure of a variety of many-particle
systems, such as liquids, glasses, crystals, heterogeneous
materials and granular media \cite{Zallen, Ashcroft, Torquato,
Edwards}. Packing problems in dimensions higher than three are
intimately related to the best way of transmitting stored data
through a noisy channel \cite{Conway}.

 A problem of
great interest is the determination of the densest arrangement(s)
of such particles and the associated density $\phi_{max}$. 
Packings of congruent circular disks in two dimensions and spheres
in three dimensions have been intensively studied.It has
been proved that the triangular lattice and face-centered cubic
lattice have the maximal packing density for circular disks
($\phi_{max}\approx 0.91$) and spheres ($\phi_{max}\approx
0.74$)\cite{Hales}, respectively. Some progress
has been made to identify good candidates for the densest
packings when the particles have a size distribution \cite{Torquato},
but primarliy in two dimensions \cite{Toth,Likos,Uche}. However, very few results are
known for the densest packings of nonspherical particles. For
ellipses ($d=2$), the densest packing is constructed by an affine
transformation of the triangular-lattice packing of circular disks
($\phi_{max}\approx 0.91$) \cite{Conway}, which can also be
obtained by enclosing each ellipse with a hexagon with minimum
area that tessellates the space \cite{Pach,Alexks}. For ellipsoids
($d=3$), the maximal known packing density ($\phi_{max}\approx
0.77$) is achieved by crystal packings of congruent ellipsoids in
which each ellipsoid has contact with 14 others \cite{Alexks}.
Recently, Conway and Torquato \cite{ConwaySal} constructed the
densest known packings of regular tetrahedra. Little is known
about the densest packings of other nonspherical particles, such as
``superballs'', as we will explain below.

\begin{figure}
\begin{center}
$\begin{array}{c@{\hspace{0.4cm}}c@{\hspace{0.4cm}}c} \\
\includegraphics[height=2.0cm, keepaspectratio]{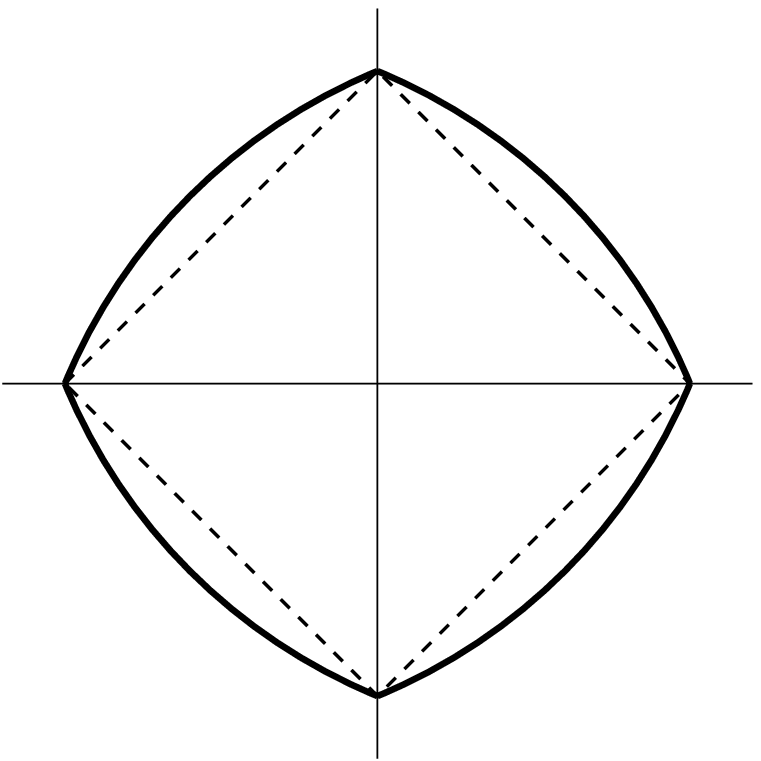} &
\includegraphics[height=2.0cm, keepaspectratio]{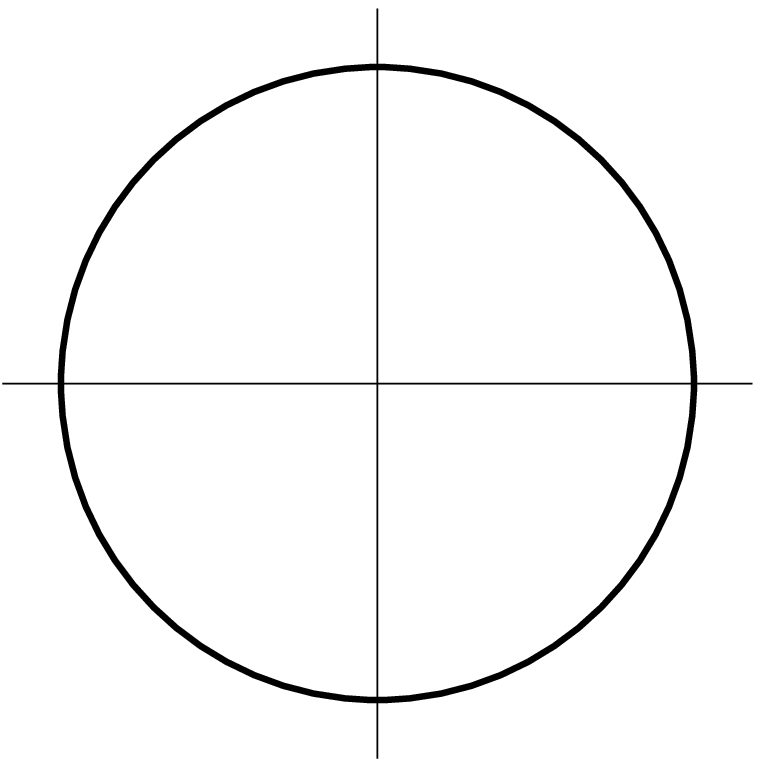} &
\includegraphics[height=2.0cm, keepaspectratio]{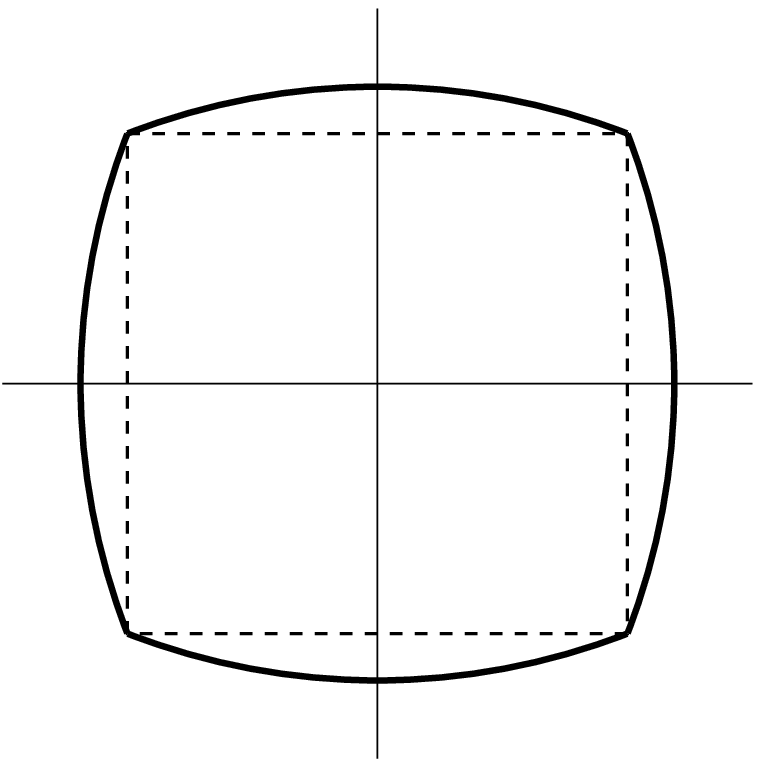} \\
\mbox{(a) p = 0.75} & \mbox{(b) p = 1.0} & \mbox{(c) p = 2.0} \\
\end{array}$
\end{center}
\caption{Superdisks with different deformation parameter $p$.}
\label{fig1}
\end{figure}

A particle is centrally symmetric if it has a center $P$ that
bisects every chord through $P$ connecting two boundary points
of the particle. A particle is convex if the entire line segment
connecting two points of the particle also belongs to the
particle. A $d$-dimensional \textit{superball} is a centrally
symmetric convex body in $d$-dimensional Euclidean space occupying
the region

\begin{equation}
\label{eq01} |x_1|^{2p}+|x_2|^{2p}+\cdots+|x_d|^{2p} \le 1,
\end{equation}

\noindent where $x_i$ $(i=1,\ldots,d)$ are Cartesian coordinates
and
 $p \ge 0.5$ is the \textit{deformation parameter}, which indicates
 to what extent the shape of the particle has deformed from an
 $d$-dimensional sphere. In particular, a \textit{superdisk} $G$ is defined by

\begin{equation}
\label{eq1} |x|^{2p}+|y|^{2p}\le1,
\end{equation}

\noindent where $(x, y)$ are Cartesian coordinates. When $p=1$,
the superdisk is just a circular disk. As $p$ continuously
increases from 1 to $\infty$, one can get a family of superdisks
with square symmetry; as $p$ decreases from 1 to 0.5, one can get
another family of superdisks still possessing square symmetry, but
the symmetry axes rotate 45 degrees with respect to that of the
first family (see Fig.~\ref{fig1}). At the limiting points $p =
\infty$ and $p = 0.5$, the superdisk becomes a perfect square.
When $p<0.5$, the superdisk becomes concave. In the following, we
simply refer to convex superdisks as ``superdisks'' for
convenience; and all concave superdisks are explicitly referred to
as ``concave superdisks'' to avoid confusion.

The study of packings of superdisks dates back to Minkowski
\cite{Minkowski}, who formulated the problem in the language of
the geometry of numbers. Thus, it is necessary for us to briefly
introduce some basic definitions and notions in the geometry of
numbers (for a general discussion of this subject, see
\cite{Cassel}). A Bravais lattice $\Lambda$ in $\Re^d$ is a subgroup 
consisting of the integer linear combinations of the vectors that constitute 
a basis for $\Re^d$. Henceforth, we will simply refer to "Bravais lattice" 
as "lattice" for convenience. A lattice packing of identical centrally 
symmetric particles is 
one in which the centers of such nonoverlapping particles are located 
at the lattice points of $\Lambda$. In a two-dimensional lattice packing 
of superdisks, the space $\Re^2$ can be geometrically 
divided into identical regions 
$F$ called fundamental cells, each of which contains the center of just 
one superdisk. An admissible lattice of a region is the one that
has no lattice points in the region except for the origin. The
critical lattice of a region is the admissible lattice whose fundamental
cell has the smallest volume. Minkowski \cite{Minkowski} conjectured that there are
two families of $G$-admissible lattices $\Lambda_0$ and
$\Lambda_1$, consistent with the symmetry of superdisk $G$, one of
which will be the critical lattice $\Lambda_c$ of $G$ for
different values of deformation parameter $p$ ($0<p<\infty$),
i.e.,

\begin{equation}
\label{eq2} \Delta_c(G) = min(\Delta_0,~ \Delta_1),
\end{equation}

\noindent where $\Delta_i$ is the volume of the fundamental cell of lattice
$\Lambda_i$ ($i=c, 0, 1$); both $\Lambda_0$ and $\Lambda_1$ have
six points on the boundary of $G$, and $(1,0)\in \Lambda_0$,
$(2^{-1/2p}, 2^{-1/2p})\in \Lambda_1$ (the lattices are defined
uniquely under these conditions). It has been shown that the
critical lattice of $G$ gives the densest lattice packing of
$\frac{1}{2}G$ \cite{Cassel}, which is defined by

\begin{equation}
\label{eq3} |x|^{2p}+|y|^{2p} \le \frac{1}{2^{2p}}.
\end{equation}

Many works that followed were devoted to this conjecture
\cite{Davis, Mordell, Cohn, WatsonI, WatsonII, GlazI, KukharevI,
KukharevII, Malyshev}. In particular, Davis obtained the proper
intervals of $p$, in which one of the families of lattices is the
critical lattice \cite{Davis}, i.e., there exists a constant
$p_0$, with $1.285 < p_0 < 1.29$, such that

\begin{equation}
\label{eq4} \Delta_c(G) = \left \{{
\begin{array}{c@{\hspace{0.1cm}}c}
\Delta_1 \quad & 0.5\le p \le1, ~ p \ge p_0, \\
\Delta_0 \quad & 1 \le p \le p_0.
\end{array}}\right .
\end{equation}

\noindent  Mordell proved the conjecture for $p=4$ \cite{Mordell};
and Malyshev et al. \cite{Malyshev} proved the conjecture for
$p\ge 6$ using a parametrization method introduced by Cohn
\cite{Cohn}. In addition, Kukharev worked out a method for
examination of Minkowski's conjecture for every specific $p$
(except for $p$ near 0.5, 1 and $p_0$) and checked the conjecture
for $p=0.65,~0.7,~0.75,~0.8,~0.85,~1.1,~1.15,~1.5,~2.0,~2.5$
\cite{KukharevI, KukharevII}. Note that the validity of the
conjecture for the trivial cases when $p = 0.5$ or $\infty$
(square) and $p = 1$ (circular disk) are well known.

Recently, Elkies et al. \cite{Elkies} obtained improvements to the
Minkowski-Hlawka bound \cite{Hlawka} on the maximal lattice-packing density for
many centrally symmetric convex bodies by generalizing the method
proposed by Rush \cite{Rush} for convex bodies symmetrical through
the coordinate hyperplanes. In particular, Elkies et al. showed
that for very large dimensions ($d\rightarrow \infty$), a small
change of $p$ from unity can give an exponential improvement on
the lower bound of the maximal lattice-packing density of
superballs.

The results in Ref.\cite{Elkies} motivates us to consider whether
or not such a dramatic improvement of packing density could
actually occur in low dimensions (e.g., $d=2,3$). Here we construct 
the densest known packings of superdisks suggested by MD simulations 
and show that one can also get a significant increase of the maximal
packing density $\phi_{max}$ of superdisks as $p$ changes from
unity, i.e., as one moves off the circular-disk point. For $p \neq
1$, the rotational symmetry of a circular disk is broken (see
Fig.~\ref{fig1}), which results in a cusp in $\phi_{max}$ at $p=1$, i.e., 
the initial increase of $\phi_{max}$ is linear in $|p-1|$. We note that the
mechanism of increasing the density of superdisk packings 
is different from that for random packings of ellipses or 
random and crystal packings of ellipsoids \cite{Alexks, AlexksIII}, 
in which a larger average number of contacts for each particle than that 
in sphere packings is required. 
The densest Bravais lattice 
packing of ellipses (ellipsoids), in which the six (twelve) contacts 
per particle is maintained, does not give an improvement on the maximal
packing density. However, one can take advantage
of the four-fold rotational symmetry of superdisks, i.e., arrange
them with proper orientations on the sites of certain lattices, to
construct packings with a dramatically improved density 
but with six contacts per particle.



We use a recently developed event-driven molecular dynamics (MD)
algorithm to generate dense (ordered and disordered) packings of
superdisks \cite{AleksII}. The MD simulation technique generalizes
the Lubachevsky-Stillinger (LS) sphere-packing algorithm
\cite{LSpacking} to the case of other centrally symmetric convex
bodies (e.g., ellipsoids and superballs). Initially, small
superdisks are randomly distributed and randomly oriented in a box
(unit cell) with periodic boundary conditions and without any
overlap. The superdisks are given translational and rotational
velocities randomly and their motion followed as they collide
elastically and also expand uniformly, while the unit cell deforms
to better accommodate the packing. After some time, a jammed state
with a diverging collision rate $\gamma$ is reached and the
density reaches a maximum value.

\begin{figure}
\begin{center}
$\begin{array}{c@{\hspace{1.5cm}}c} \\
\includegraphics[height=5.0cm, keepaspectratio]{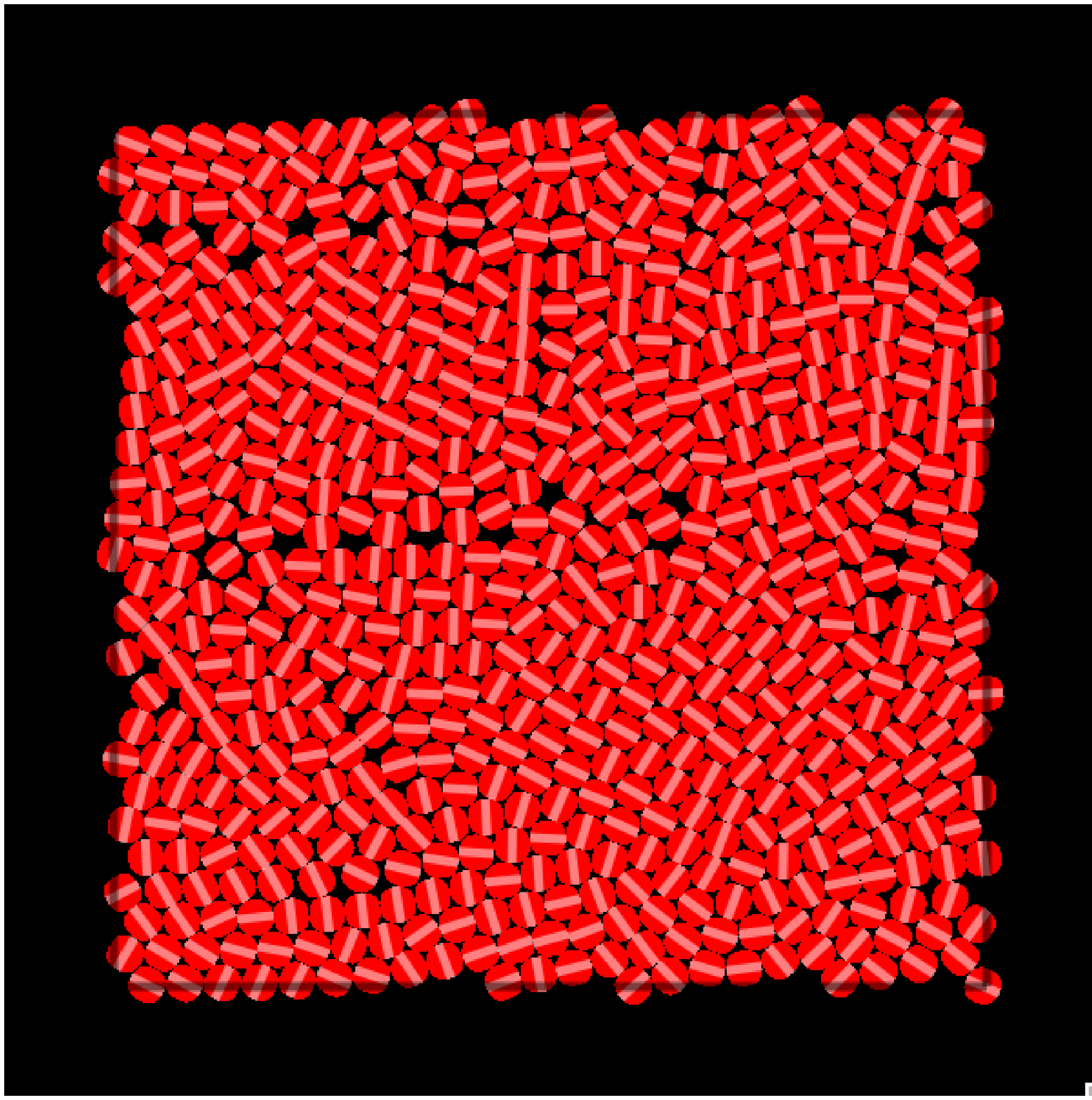} &
\includegraphics[height=5.0cm, keepaspectratio]{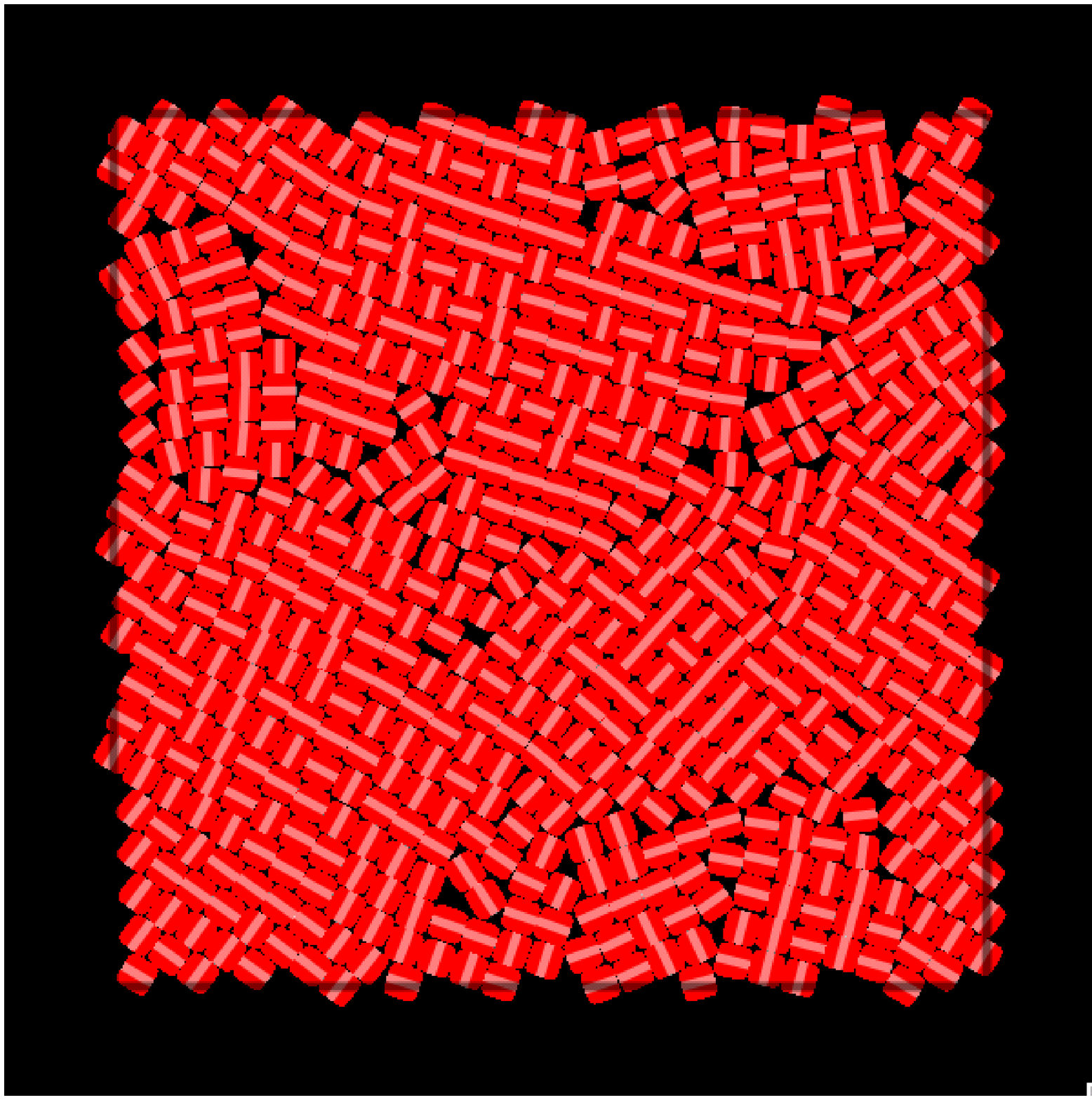} \\
\mbox{(a) p = 1.25} & \mbox{(b) p = 2.5} \\
\end{array}$
\end{center}
\caption{Large packings of superdisks with different deformation
parameters generated by MD simulation. The number of particles in
the unit cell is $N = 625$. The white ``chord'' in each superdisk
indicates  one of its symmetry axes. The boundaries of the simulation box are shown by  dark lines.} \label{fig2}
\end{figure}

Our aim is to produce dense packings of superdisks and
to identify the densest packing structures if possible. Extensive
experience with spheres and circular disks has shown that, for reasonable
large packings, sufficiently slowing down the growth of the
density, so that the hard-particle system remains close to the
equilibrium solid branch of the equation of state, leads to
packings near the face-centered-cubic lattice and triangular
lattice, respectively \cite{SalPRL, Kansal}. However, this
requires impractically long simulation times for large superdisk
packings. We note that in two dimensions, because the densest
local packing of circular disks (a triangle with three circular disks centered at
its corners) can tessellate the space, large packings of circular disks are
usually nearly completely crystallized, i.e., they contain grains
of circular disks on triangular lattice and dislocations, even when a moderate
expansion rate is used. We find from simulations that this is also
true for packings of superdisks (see Fig.~\ref{fig2}), which
implies the densest equilibrium state (densest packing) of
superdisks is consistent with the structure of the densest local
clusters that tessellate space. Thus, we should be able to
identify the \textit{possible} densest packings of superdisks by
running the simulation for unit cells with a small number of
particles (e.g., from 4 to 16 particles).

\begin{figure}
$\begin{array}{c@{\hspace{1.5cm}}c} \\
\includegraphics[height=5.0cm, keepaspectratio]{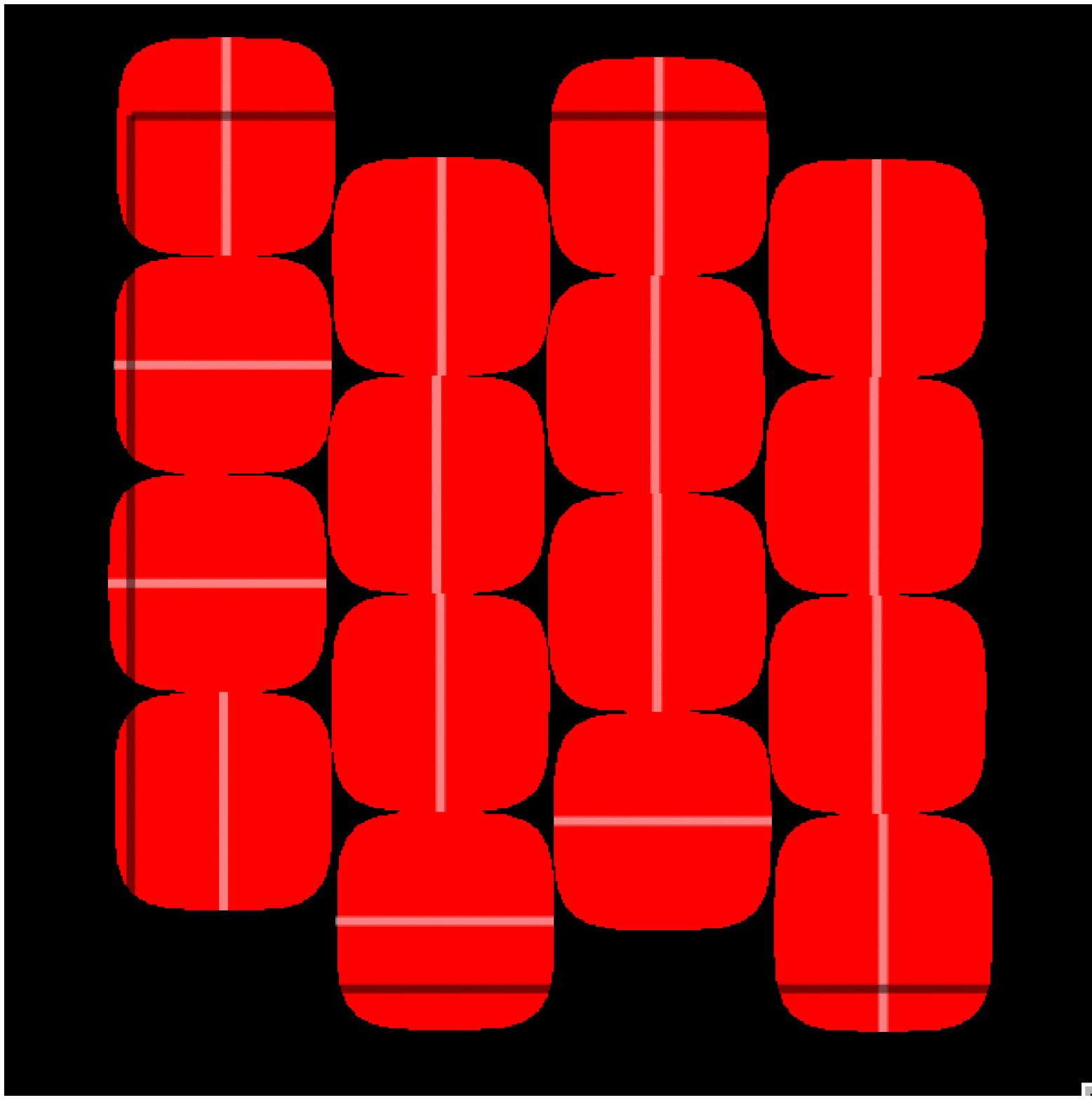} &
\includegraphics[height=5.0cm, keepaspectratio]{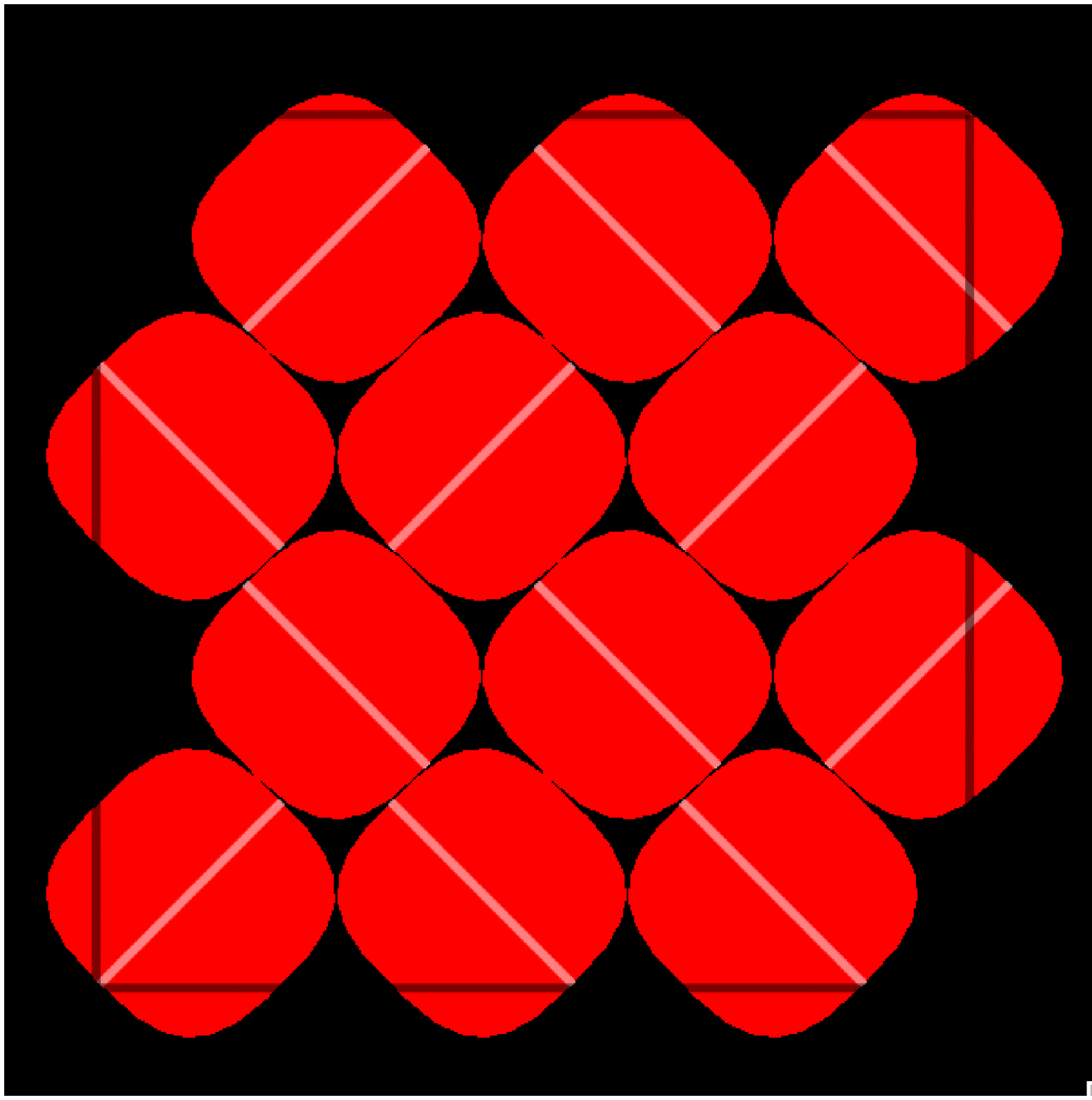} \\
\mbox{(a) $\Lambda_0$-lattice packing} & \mbox{(b) $\Lambda_1$-lattice packing} \\
\end{array}$
\caption{Two types of lattice packings of superdisks that have the
maximal packing density for different deformation parameter $p$.
In the figures (a) and (b), $p = 2.0$ and $p = 1.5$, respectively.
In both cases $\Lambda_1$-packing is denser. The
white ``chord'' in each superdisk indicates one of its symmetry
axes. The boundaries of the simulation box are shown by  dark lines.} \label{fig3}
\end{figure}

\begin{figure}
$\begin{array}{c}\\
\includegraphics[height = 5.5cm, keepaspectratio]{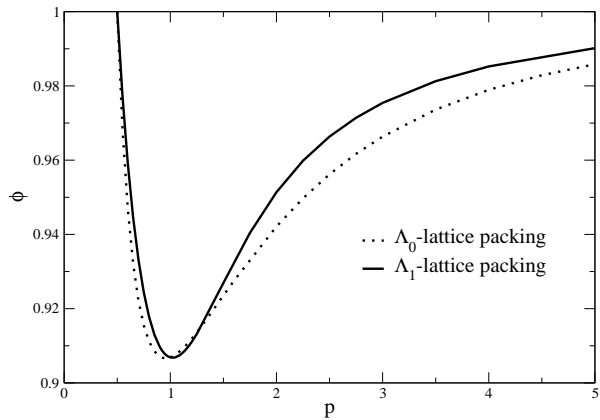}\\
\end{array}$
\caption{Densities of $\Lambda_0$-lattice and $\Lambda_1$-lattice
packings of superdisks as a function of deformation parameter
$p$.} \label{fig4}
\end{figure}

Two types of lattice packings of superdisks, which gives the
maximal packing density among all packings generated by
simulations for different values of deformation parameter $p$, are
shown in Fig.~\ref{fig3}. Note that we do not exclude the
possibility of the existence of denser periodic packings with
complex basis, although we didn't find any of these packings by
examining the dense grains of large packings of superdisks, which
would contain denser clusters if they exist. Given that the superdisks
are defined by Eq.~(\ref{eq3}), it is clear that the two types of
lattices we found are $G$-admissible lattices $\Lambda_0$ and
$\Lambda_1$, respectively. Subsequent analytical calculations
suggested by the simulation results gives us the packing densities
as a function of $p$ (see Fig.~\ref{fig4}) as well as the value of
$p_0$, i.e., $p_0 \approx 1.2863$. As can be seen from
Fig.~\ref{fig4}, the packing density increases dramatically as $p$
moves away from the circular-disk point ($p=1$). It is worth 
noting that our MD algorithm can be
employed to verify Minkowski's conjecture for all values of $p\ge
1$ to a very high numerical accuracy in principle.

As the deformation parameter $p$ changes from unity, the
rotational symmetry of circular disks is broken, i.e., superdisks only
possess four-fold rotational symmetry. Donev et al. studied the
effects of broken symmetry introduced by stretching circular disks
(spheres) into ellipses (ellipsoids) on the random packings of
these particles \cite{AlexksIII, AlexksIV}, e.g., how the average
contact number and packing density change as a function of the
maximum aspect ratio of the ellipses and ellipsoids. Donev et al.
also constructed a family of unusually dense crystal packings of
ellipsoids, by taking advantage of the broken symmetry via the new
rotational degrees of freedom that result for nonspherical
particles \cite{Alexks}. Densest packing of ellipses can be
trivially obtained by an affine transformation of
triangular-lattice packing of circular disks, which produces a lattice
packing of ellipses with the packing density unchanged (i.e.,
$\phi_{max} \approx 0.91$). As stated earlier, 
this can be shown by enclosing each ellipses
with a hexagon with the smallest area which tessellates the space
\cite{Pach}.

The broken symmetry of superdisks, introduced by deforming the
circular disks, affects the packing density in a non-trivial way. In
particular, we focus on the maximal packing density $\phi_{max}$,
which is a property of equilibrium superdisk system in solid
state, as a function of $p$. There are two discontinuities of the
derivative $\phi'_{max}(p)$ at $p=1$ and $p=p_0$, respectively
(see Fig.~\ref{fig4}). Thus, as $p$ changes from 1, $\phi_{max}$
will increase in a cusp-like manner. By expanding $\phi_{max}(p)$
around $p=1$, we get

\begin{equation}
\label{eq6} \phi_{max} = \phi_0[1-0.009(p-1)+O((p-1)^2)]
\end{equation}

\noindent for $0.5\le p\le 1$ and

\begin{equation}
\label{eq7} \phi_{max} = \phi_0[1+0.01(p-1)+O((p-1)^2)]
\end{equation}

\noindent for $p\ge1$, where $\phi_0 = 0.906...$ is the density of
triangular-lattice packing of circular disks. In other words, the initial increase
in the density is linear in $|p-1|$, which is analogous to the 
effects of asphericity of ellipses (ellipsoids) on
the density of random packings of these particles \cite{AlexksIV}.
The increase of the density of packings of ellipses
(ellipsoids) is related to the increase of the average number of 
contacts per particle. The Bravais lattice packing of ellipses, 
which can be obtained by an affine transformation of the triangular-lattice 
packing of circular disks, does not have an improvement on the maximal 
packing density since the six contacts per particle is maintained. 
Here the increase of the maximal
packing density of superdisks with certain $p \neq 1$ is due to
the fact that the four-fold rotationally symmetric shape of
superdisk is more efficient for filling space, which we will
discuss in detail in the following.

The other discontinuity at $p = p_0$ corresponds to a jump-like
change of the packing structure that evolves as $p$ varies. In
particular, as $p$ increases from 1 ($\Lambda_0$-lattice and
$\Lambda_1$-lattice coincide at $p=1$), the packing lattice
continuously deforms from triangular-lattice to
$\Lambda_0$-lattice till $p=p_0$, where the packing lattice
``jumps'' from $\Lambda_0$-lattice to $\Lambda_1$-lattice and then
proceeds to deform continuously. This is because the superdisk
fits the enclosing cell (defined in the following) of $\Lambda_1$-lattice
better when $p$ exceeds $p_0$.


\begin{figure}
\begin{center}
$\begin{array}{c@{\hspace{0.8cm}}c} \\
\includegraphics[height=3.5cm, keepaspectratio]{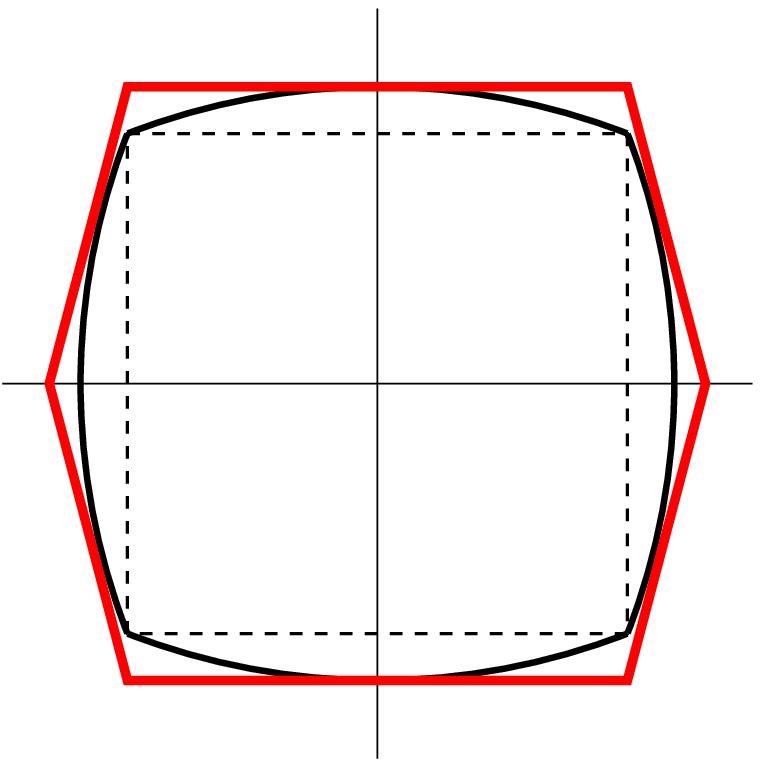} &
\includegraphics[height=3.5cm, keepaspectratio]{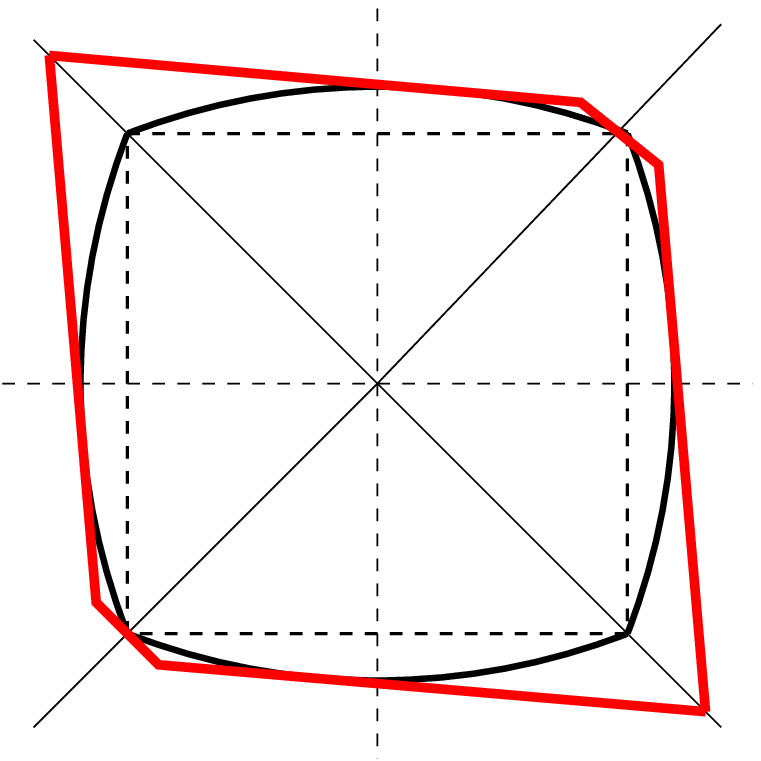} \\
\mbox{(a) $C_0$ for $\Lambda_0$-lattice} & \mbox{(b) $C_1$ for $\Lambda_1$-lattice} \\
\end{array}$
\end{center}
\caption{The enclosing cells $C_0$ and $C_1$ of superdisk with p =
2.0 for $\Lambda_0$-lattice and $\Lambda_1$-lattice,
respectively.} \label{fig5}
\end{figure}

Analysis of the packing structure is necessary for understanding
the aforementioned effects of broken symmetry. We define the
\textit{enclosing cell} $C$ of a superdisk to be the polygon whose
edges are common tangent lines of the superdisk and its contact
neighbors (see Fig.~\ref{fig5}). For a particular lattice packing,
the enclosing cells for all superdisks are the same; this
enclosing cell must be able to tessellate space. As $p$ varies,
the enclosing cell for a particular lattice also deforms
continuously, e.g., from hexagon to square as $p$ increases from 1
to $\infty$.

For a fixed value of $p$, the denser lattice packing is the one
with the smaller enclosing cell (i.e., the enclosing cell that fits
the superdisk better). As can be seen in Fig.~\ref{fig5}, the two
types of enclosing cells $C_0$ and $C_1$ (associated with
$\Lambda_0$-lattice and $\Lambda_1$-lattice, respectively)
accommodate the curvature around the boundary point $(2^{-1/2p},
2^{-1/2p})$ and its images, and $(1,0)$ and its images better,
respectively, to give a higher local density. When $p$ is slightly
larger than 1, the curvature around point $(2^{-1/2p}, 2^{-1/2p})$
and its images is dominant, and the denser packing is given by
$\Lambda_0$-lattice. As $p$ increases, the curvature around point
$(1,0)$ and its images becomes dominant, thus, the packing jumps
to $\Lambda_1$-lattice. For $0.5\le p\le 1$, the curvature around
point $(1,0)$ is always dominant, so the $\Lambda_1$-lattice gives the
denser packing for all $p$ in that interval.

\begin{figure}
\begin{center}
$\begin{array}{c@{\hspace{0.5cm}}c} \\
\includegraphics[width=1.75cm, keepaspectratio]{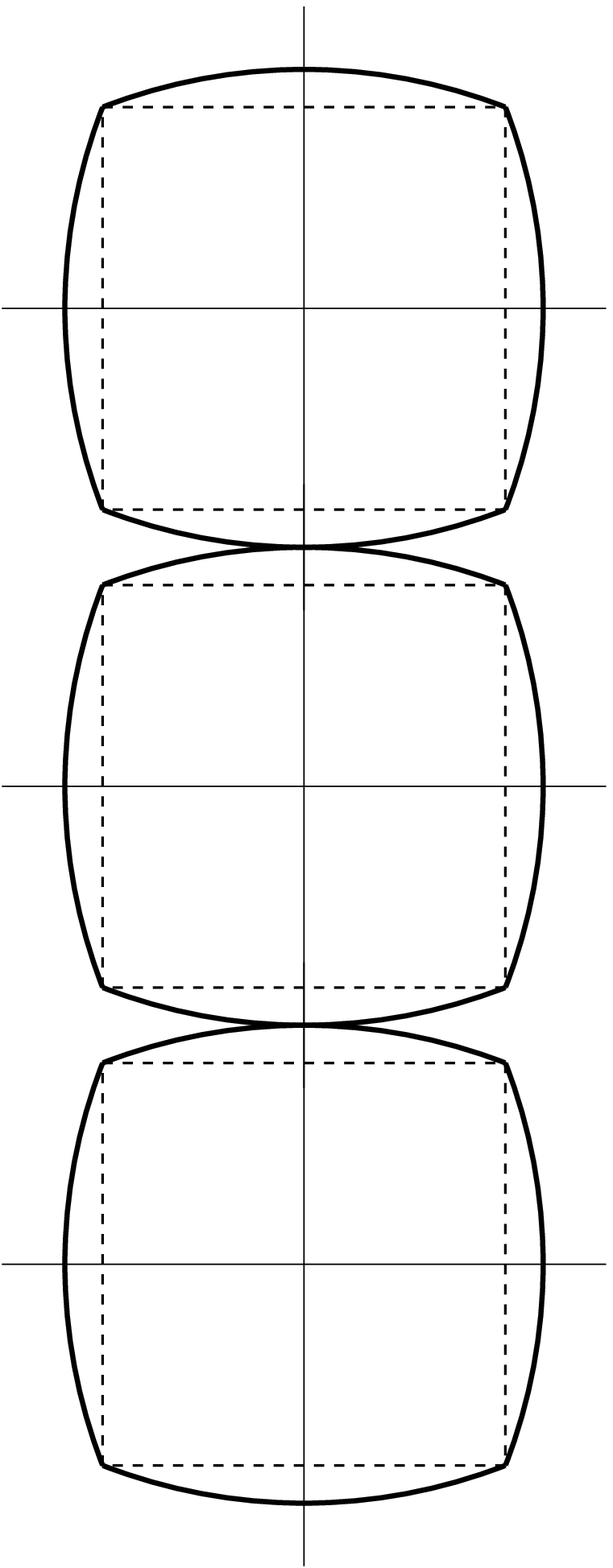} &
\includegraphics[height=1.75cm, keepaspectratio]{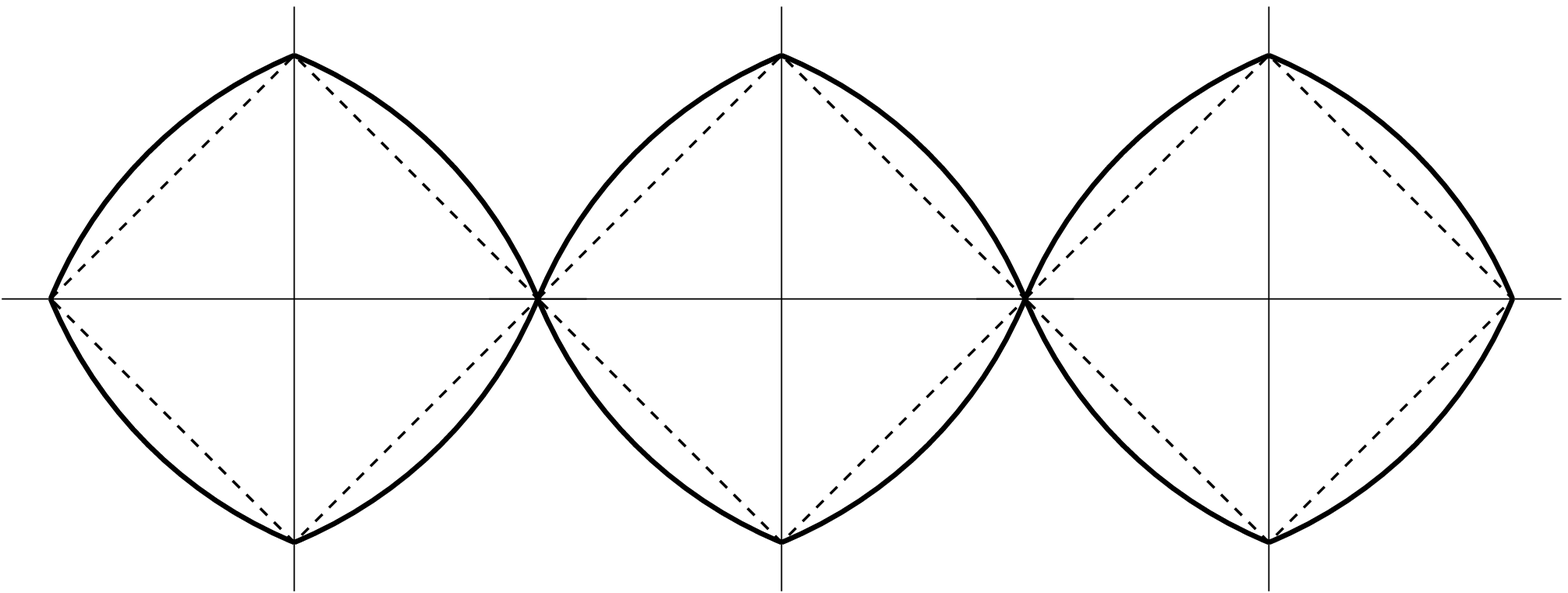} \\
\mbox{(a) p = 2.0} & \mbox{(b) p = 1.5} \\
\end{array}$
\end{center}
\caption{Two types of superdisk chains, as the building block for
lattice packings of superdisks shown in Fig.~\ref{fig3}.}
\label{fig6}
\end{figure}

It is also of interest to consider the packings as stacks of
superdisk chains, as shown in Fig.~\ref{fig6}. For example, the
lattice packings in Fig.~\ref{fig3}(a) and (b) can be constructed
by stacking the superdisk chains shown in Fig.~\ref{fig6}(a) and
(b) horizontally and vertically, respectively. This may seem to be
trivial for superdisks, however this view of packing structure
enables us to construct dense packings of concave particles and to
better understand packings of superballs in higher dimensions.


\begin{figure}
\begin{center}
$\begin{array}{c@{\hspace{0.8cm}}c} \\
\includegraphics[height=3.5cm, keepaspectratio]{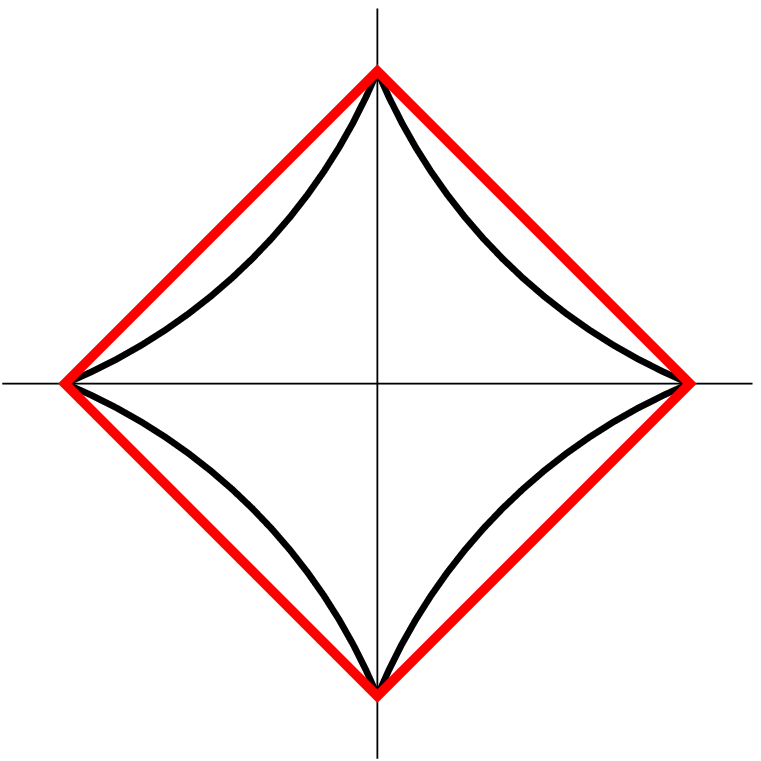} &
\includegraphics[height=3.5cm, keepaspectratio]{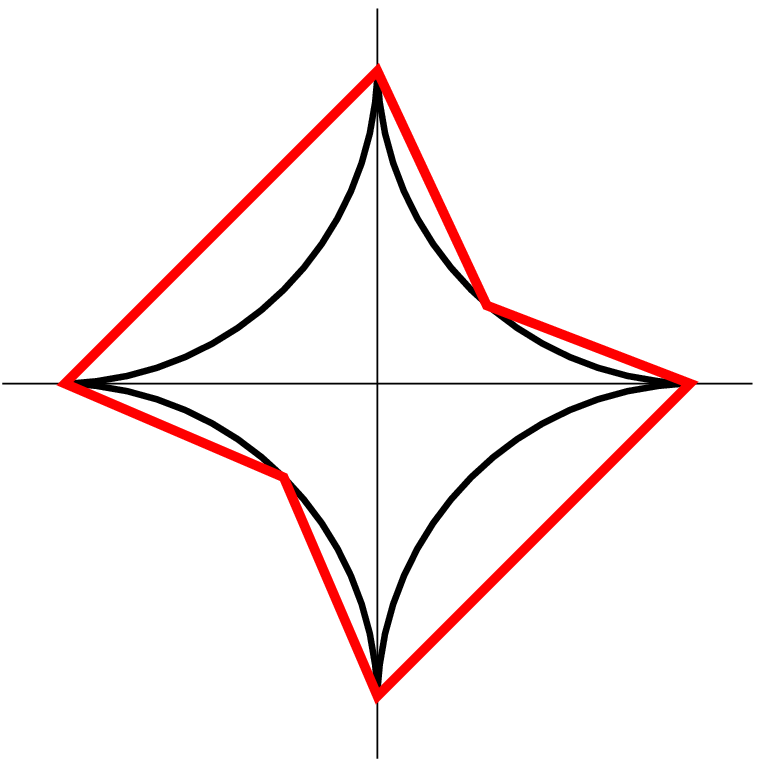} \\
\mbox{(a) p = 0.48} & \mbox{(b) p = 0.45} \\
\end{array}$
\end{center}
\caption{(a) A concave superdisk with convex enclosing box (a
square). (b) A concave superdisk with concave enclosing box (an
hour-glass).} \label{fig7}
\end{figure}

\begin{figure}
$\begin{array}{c@{\hspace{0.5cm}}c} \\
\includegraphics[height=3.8cm, keepaspectratio]{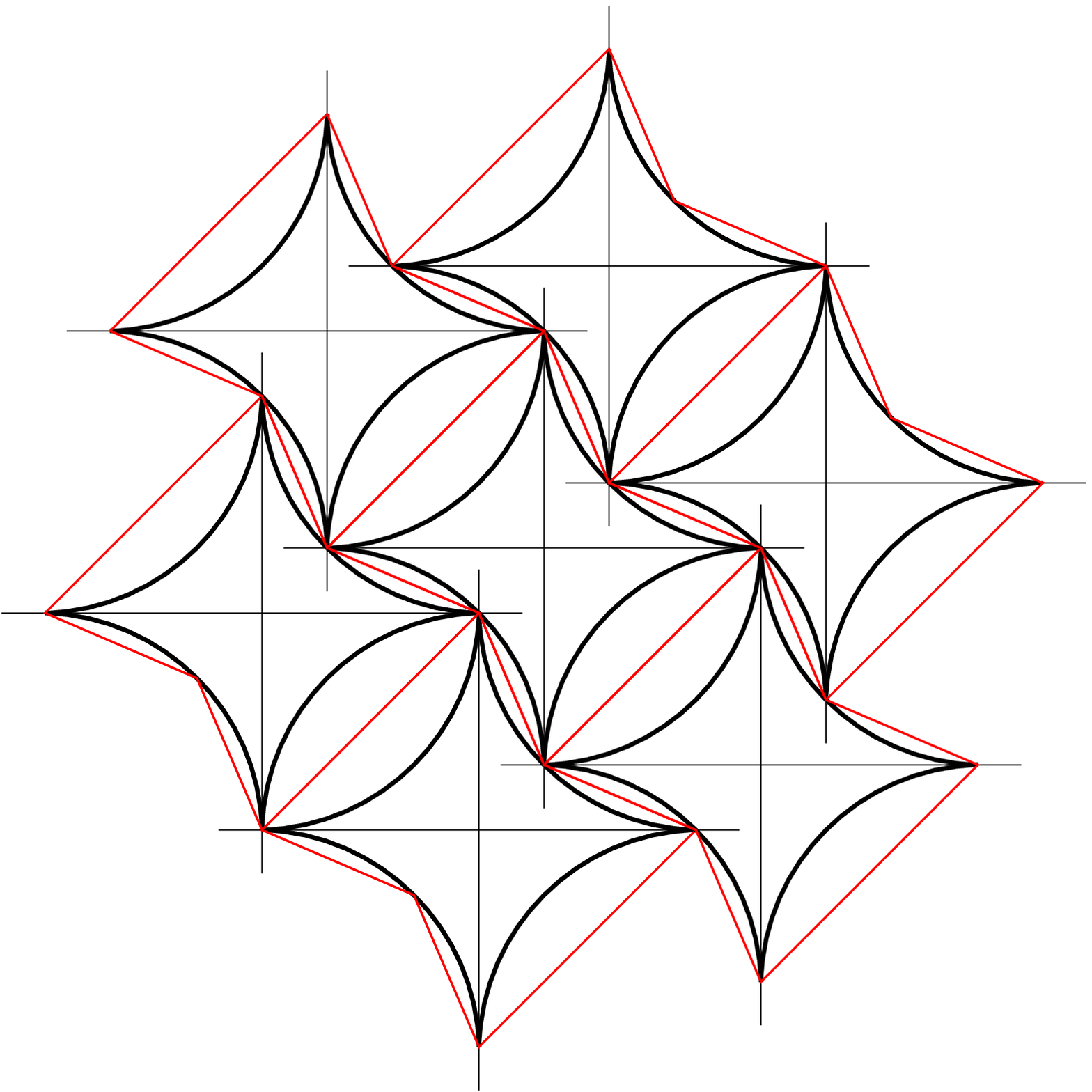} &
\includegraphics[height=3.8cm, keepaspectratio]{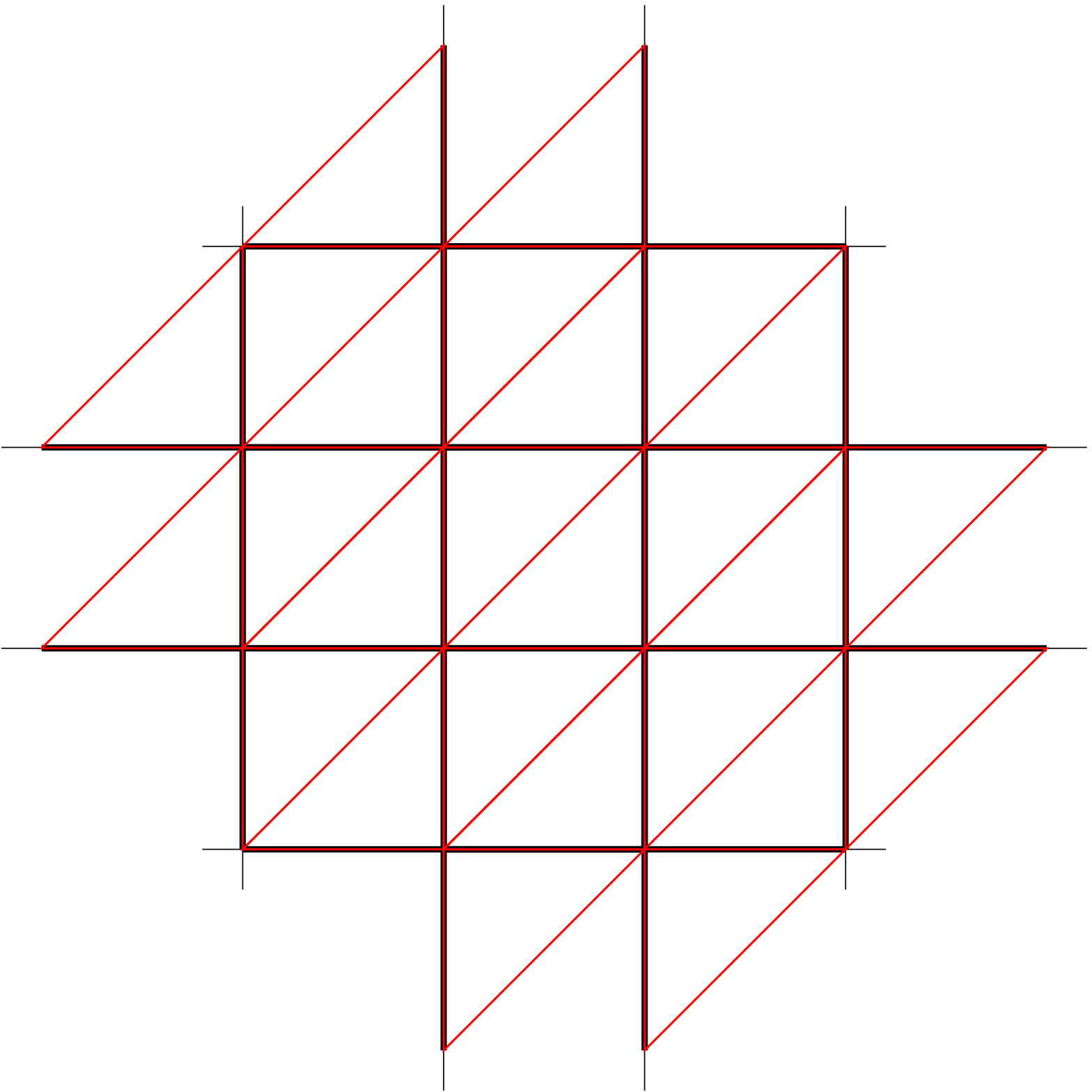} \\
\mbox{(a) p = 0.45} & \mbox{(b) p =  0} \\
\end{array}$
\caption{Examples of dense packings of concave superdisks
constructed by using the method described in the text.} \label{fig8}
\end{figure}

\begin{figure}
$\begin{array}{c}\\
\includegraphics[height = 5.5cm, keepaspectratio]{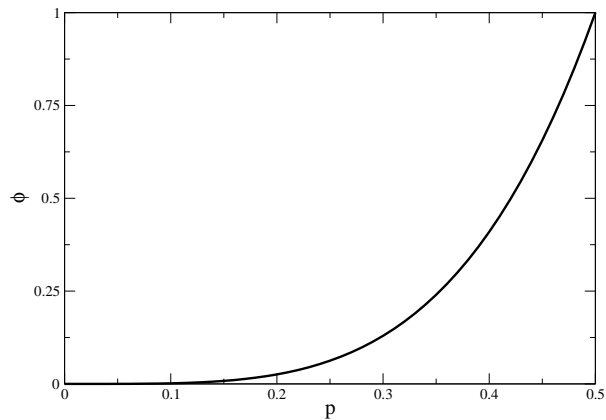}\\
\end{array}$
\caption{Densities of the constructed lattice packings of concave
superdisks as a function of deformation parameter $p$.}
\label{fig9}
\end{figure}

It is interesting to generalize the above discussion to the case
of concave superdisks ($0<p< 0.5$), as shown in
Fig.~\ref{fig7}. To construct dense packings of these particles,
one needs to take advantage of the concave shape to reduce
exclusion-volume effects. (The exclusion volume of a particle is
the region around its center in which no other particle centers
can be found due to the impenetrability constraint.)

For each concave superdisk, we can define a convex enclosing box
as the one has the smallest area among all convex boxes that
contain the particle, which is a square in this case [see
Fig.~\ref{fig7}(a)]. First, we construct densest packing of the
convex enclosing boxes, i.e., stacks of square chains. Then, we
allow these square chains to overlap as much as possible without
violating the impenetrability constraints imposed by the hard
particles. This also maximizes the number of contact neighbors for
every concave superdisk. In this way, we can construct a family of
dense lattice packings of concave superdisks, in which each
particle has an hour-glass-like concave enclosing box [see
Fig.~\ref{fig7}(b)] and 6 contact neighbors. Note that the
constructed lattice packings of concave superdisks resemble
$\Lambda_0$-lattice packings of superdisks.

Two examples of the dense lattice packings of concave superdisks
constructed by using the aforementioned method are shown in
Fig.~\ref{fig8}. As $p$ decreases from 0.5, the particle shrinks;
at the limit $p\rightarrow 0$, the superdisks become ``crosses''
[see Fig.~\ref{fig8}(b)]. Note that at the limiting case of
``crosses'', the area of the particle is 0, so is the packing
density defined as the area fraction of the space covered by the
particles. However, the \textit{number density} of the lattice
packing we constructed is \textit{twice} of that for
square-lattice packing of crosses, whose enclosing box is a
square. The density of the constructed lattice packings of concave
superdisks as a function of $p$ is shown in Fig.~\ref{fig9}. We
emphasize that the existence of denser periodic packings or other
lattice packings of these concave particles is also possible.


In summary, we have constructed the densest known packings of 
superdisks suggested by 
MD simulations. In particular, we found that the
maximal packing densities of superdisks for certain $p\neq 1$ are
achieved by one of the two families of lattice-packings, i.e.,
$\Lambda_0$-lattice and $\Lambda_1$-lattice packings, which
provides additional numerical evidence for Minkowski's conjecture
concerning the critical determinant of the region occupied by a
superdisk. We also showed that the increase of maximal packing
density is initially linear in $|p-1|$; $\phi_{max}(p)$ has a cusp at $p=1$
and has another discontinuity in its derivative $\phi'_{max}(p)$
at $p = p_0$, which are effects of the broken symmetry of
superdisks. The result that $\phi_{max}$ increases as $p$ varies
from 1 is also consistent with the improvement on the lower bound
on the packing density of superballs in arbitrarily high dimensions
found by Elkies et al. Based on our observations of the
structural properties of lattice packings of superdisks, we also
proposed an analytical method to construct dense packings of concave
superdisks. We emphasize that a good understanding of packings of
superdisks provides the basis for the study of packings of
superballs and superellipsoids in higher dimensions.

In future work, we will generate and study both ordered and
disordered packings of superballs and superellipsoids in three
dimensions, focusing on the role of broken symmetry in influencing
the properties of such packings. It is important to adapt the 
methodology described in Ref.~\cite{AlexksIV} to test the jamming categories
\cite{SalJPC} of disordered packings of superballs and
superellipsoids in order to study of the maximally random jammed
(MRJ) states of packings with such nonspherical particle shapes.

\begin{acknowledgments}
The authors would like to thank Aleksander Donev for his valuable
suggestions on the MD packing algorithm. Y.J. and S.T. were
supported in part by the National Science Foundation under Grant
No. DMS-0312067.
\end{acknowledgments}

\end{document}